\documentclass[twocolumn,showpacs,prl,aps]{revtex4}
\usepackage{epsf}

\begin{document}
\title{Storage of localized structure matrixes in nematic liquid crystals}

\author{Umberto Bortolozzo and Stefania Residori} 

\address{Institut Non Lin\'eaire de Nice Ð 1361 Route des Lucioles Ð 06560 Valbonne - Sophia Antipolis Ð France}

\date{\today}

\begin{abstract}
We  show experimentally that  large matrixes of localized structures can be stored as elementary pixels in a nematic liquid crystal cell. Based on optical feedback with phase modulated input beam, our system allows to store, erase and actualize in parallel the localized structures in the matrix. 
\end{abstract}


\maketitle

Non-equilibrium processes lead in nature to the formation of spatial patterns, sometimes appearing in a restricted region of the space, so that we deal with localized instead of extended structures \cite{Cross}. During the last years, localized structures have been observed in different fields, such as in chemical reactions \cite{1}, granular media \cite{2}, plasmas \cite{3}, surface waves \cite{4} and, in optics \cite{Newell}, in photorefractive crystals \cite{5}, liquid crystals \cite{6}, atomic vapors \cite{7} and semiconductor micro-cavities \cite{8}. 
Localized states have attracted much attention in view of their potential applications as elementary pixels for information storage and retrieval \cite{Firth,Tlidi}. 
In particular, parallelism is nowadays an important point to be addressed in liquid crystal technology, thus optical localized structures in non pixellized liquid crystal devices, such as the Liquid-Crystal-Light-Valve (LCLV) \cite{Residori},
constitute a great advancement towards the realization 
of a system able to respond simultaneously at multi-distributed spatially continuous inputs.  A fundamental feature of optical localized structures
is their bistable behavior, which allows to switch them on and off by 
sending an appropriate (and small) perturbation.
This property marks a clear distinction from other types of localized states previously observed in nematic liquid crystals, such as the "worms" appearing in electroconvection \cite{9} and the spatial solitons propagating in bulk cells \cite{10}. However, up to now practical applications have been limited by the strong crosstalk among adjacent localized structures and by the influence of phase/intensity gradients on their stability. 

Here, we report the results of experimental investigation on phase controlled localized structures. We present a new optical scheme, based on a LCLV with optical feedback, where a computer-interfaced display stabilizes large matrixes of localized structures. We show that localized structures can be stored, erased and actualized in parallel on the matrix.

\begin{figure} [h!]
\centerline{\epsfxsize=7 truecm \epsffile{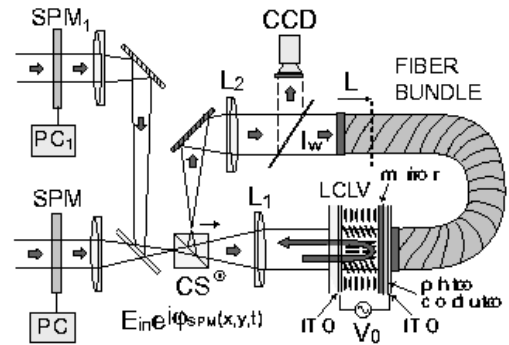}}
\caption{Experimental setup: the LCLV with optical feedback and phase modulated input beam.
\label{fig1}}
\end{figure}

As schematically depicted in Fig.\ref{fig1}, the LCLV is composed of a nematic liquid crystal \cite{11} in between a glass and a photoconductive plate over which a dielectric mirror is deposed. The nematic order is defined by an average orientation of the liquid crystal molecules, identified by an unit vector, $\vec n$, so called the director. The liquid crystals are planar aligned ($\vec n$ parallel to the cell walls) and the cell thickness is $15$ $\mu m$. Transparent electrodes covering the glass plates permit the application of an external voltage $V_0$ across the liquid crystals. The photoconductor behaves like a variable impedance, its resistance decreasing when increasing the intensity of the light $I_w$ impinging on the rear side of the LCLV( write light).
Thus, the voltage $V_{LC}$ that effectively drops across the liquid crystals is $V_{LC}= \Gamma V_0 + \alpha I_w$, where $V_0$ is the a.c. voltage externally applied to the LCLV and $\Gamma$, $\alpha$ are phenomenological parameters summarizing, in the linear approximation, the response of the photoconductor ($\Gamma$ is the dark transfer factor) \cite{Aubourg}. 

Under application of the voltage, liquid crystal molecules reorient towards the direction of the electric field \cite{11}. Because of liquid crystal birefringence, the molecular reorientation induces a refractive index change for the light incoming on the LCLV. As a result, the input beam, which pass through the liquid crystal layer and is reflected by the mirror of the LCLV, undergoes a phase shift $\varphi$, which depends on the write intensity $I_w$ and on the liquid crystal birefringence $\Delta n=n_e-n_0$, $n_e$ and $n_o$ being, respectively, the extraordinary (parallel to $\vec n$) and ordinary (perpendicular to $\vec n$) refractive index. 
For $I_w\le 2 mW/cm^2$ the phase shift is proportional to the light intensity and the LCLV behaves like a Kerr nonlinearity.
If  $\theta$  is the average reorientation angle of the molecules, and for $\Delta n \ll n_e,n_0$ we can express the phase shift as $\varphi = \beta  \cos^2 \theta$, with $\beta=2 \pi \Delta n d / \lambda$, where $\lambda$ is the optical wavelength. In our experiments $\lambda=632.8$ $nm$ and  $\Delta n=0.2$, with $n_e=1.7$ and $n_0=1.5$.

The optical feedback is obtained by sending back onto the photoconductor the light that has passed through the liquid-crystals and has been reflected by the mirror of the LCLV. In this case, the phase shift $\varphi$ experienced by the light beam depends on the liquid crystal reorientation angle, which, on its turn, induces a change of the effective voltage applied to the liquid crystals. Two lenses, $L_1$ and $L_2$, form an image of the front side of the LCLV on the plane marked by a dashed line. An optical fiber bundle closes the loop, transporting the image from one end to the other with negligible losses and with a spatial resolution of $20$ $\mu m$. The free propagation length is $L=80$ $mm$, over which diffraction takes place. For such a diffractive feedback, the system is known to display a transverse spatial instability \cite{Residori}, developing hexagonal patterns at a typical wavelength $\sqrt{2 \lambda L}$ \cite{Dalessandro}. 

In the present experiment, the beam splitter (CS) at the entrance of the optical loop is a polarizing cube, so that it transmits the vertical polarization and reflects the horizontal one. The liquid crystal director is at $45^\circ$ with respect to the input polarization, which is vertical. Since the beam reflected by the cube is horizontally polarized, the feedback loop produces polarization interference between the ordinary (polarized orthogonally to $\vec n$) and extraordinary (polarized parallel to $\vec n$) waves. This condition ensures multistability between differently oriented states of the liquid crystals and leads to the appearance of stable localized structures \cite{6,16,19}.

As schematically depicted in Fig.\ref{fig1}, the setup includes also a spatial phase modulator (SPM) connected to a personal computer (PC) and inserted 
in the optical path of the input beam. 
A second beam is represented in the figure, which pass through another spatial phase modulator, SPM$_1$ connected to PC$_1$, to stress the fact that, in principle, it is possible to couple in a parallel way two or more inputs onto the system.
However, in the experiment we have used only the first SPM
and we have sent the control inputs either sequentially or overlapped in a single image. Note that phase control requires coherence of the light beam, whereas
intensity perturbation can be sent even through incoherent control channels, so that intensity control can be achieved even by using independent light sources.

The SPM is a twisted nematic liquid crystal display without polarizers. lt may be shown in general that any polarized beam of light propagating in a such a display can be decomposed into a linear combination
of two normal modes of propagation, designed as twisted extraordinary 
($te$ mode) and twisted ordinary ($to$ mode) \cite{22}. Normal modes 
are in general elliptically polarized but in the regime of slow twist the ellipses are almost linear with their major axes following the local orientation of the liquid crystal director. The polarization of the beam impinging on the SPM
is parallel to the entrance extraordinary axis, which is at $45^\circ$ with respect to the vertical polarization. For such a geometry,
the SPM induces on the beam impinging on the LCLV
a phase shift $\varphi_{SPM}$ that is a function of the gray level set on the PC. 
We have measured $\varphi_{SPM}$ varying from $0$ to $\pi$ when the
gray level changes from $0$ to $255$.

\begin{figure} [h!]
\centerline{\epsfxsize=7 truecm \epsffile{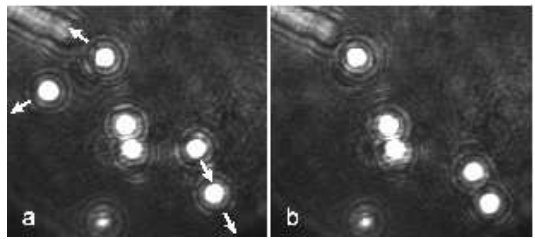}}
\caption{Localized structures without control: a) arrows indicate the drift; b) after a few seconds, one structure has exited the frame and two have formed a bound state; the bound state in the middle remains stable. 
\label{free}}
\end{figure}

As a first test, we have used the SPM to address a local pulse on the LCLV, so that we could switch on/off a single localized structure in any arbitrary position. The local pulse was either a bright or a dark spot onto the SPM, so that a localized structure was created
or deleted, respectively.
Once created, localized structures drift by following any phase/intensity gradients due to liquid crystal inhomogeneities. Moreover, adjacent localized structures interact each other, eventually forming bound states, that are very robust against external perturbation, or annihilating. These interactions
are the origin of the crosstalk between neighboring localized structures \cite{7,16}.
Such a dynamics and the formation of bound states are shown in Fig.\ref{free}. Here, the size of each localized structure was $420\pm30$ $\mu m$ and the voltage applied to the LCLV was fixed to $V_0=12.8$ $V$, frequency $6$ $KHz$.

It has been numerically demonstrated for a Kerr-like system
that localized structures behave like single particles moving in the presence of 
phase/intensity gradients \cite{Firth}. In the case of the LCLV, numerical simulations have shown that phase gradients are more efficient in displacing localized structures than intensity gradients are \cite{21}. By using this property, 
it has been demonstrated that a phase grid is able to pin localized structures on the local maxima. 

\begin{figure} [h!]
\centerline{\epsfxsize=7 truecm \epsffile{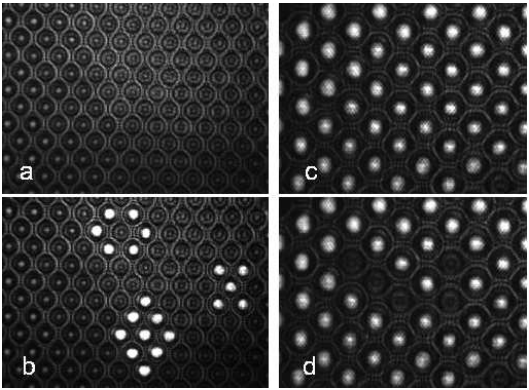}}
\caption{ Writing/erasing of localized structures : a) empty matrix; b) an arbitrary configuration of localized structures is written by flashing an image through the SPM; c) a matrix with all the pixels in the on-state; d) some pixels are switched off by a local pulse sent through the SPM. 
\label{fig4}}
\end{figure}

In order to suppress the crosstalk in the experiment, we have sent through the SPM an egg-box profile that modulates the phase of the input beam. Thus, after the cube splitter, the input beam is phase modulated with a periodic two-dimensional grid 
$\varphi_{SPM}=\varepsilon(cos K x+cos K y)^2$, where $\varepsilon = 0.5$ $rad$ and $K= 0.015$ $rad/\mu m$.  When traveling in the optical feedback loop, the beam
undergoes diffraction so that the initial phase modulation is converted into an intensity modulation \cite{Dalessandro} and phase maxima gives rise to low amplitude intensity maxima on the photoconductor. A near-field image showing the intensity modulation on the LCLV
is shown in Fig.\ref{fig4}a. This low amplitude pattern which is superimposed to the 
normally uniform background acts as a matrix where to store localized structures. If we use the same procedure as before to write
localized structures, we can see that once created
they move towards the closest local maximum of the intensity and remain attached there. 
Indeed, the energy barrier created by the phase profile is sufficiently high to fix the localized structures at the maxima locations and to keep them stable against perturbations.  
By sending through the SPM an image containing the information to be stored, we can write any arbitrary configuration of localized structures, as shown in Fig.\ref{fig4}b. 
In an analogous way, we can prepare the system in a state where all the pixels in the
matrix are switched on, as shown in Fig.\ref{fig4}c, and then erase the wanted sites by sending a local pulse through the SPM, as depicted in Fig.\ref{fig4}d. 

The period of the spatial grid is chosen in order to match the size of the
localized structures, therefore we can bring them as close one to the other as the maximum packing limit.
The parameter $\varepsilon$ ranges from $0.2$ to $0.6$ $rad$. The lower limit is dictated by the 
minimum modulation amplitude capable to overcome the cross-talk between localized structures, whereas the maximum limit has not to exceed the value for which
the homogeneous stationary state becomes unstable. 
We have set $\varepsilon=0.5$, so that the maximum 
intensity of the grid is $14\%$ of the localized structures peak intensity and
localized structures are strongly pinned. Without the phase grid, the range of existence of localized structures goes from $I_{in}=0.38 \pm 0.04$ to $0.52 \pm 0.05$ $mW/cm^2$.  For higher input intensity, we observe spontaneous formation of hexagons with spatial wavelength $205\pm 30$ $\mu m$ and spontaneous nucleation of localized peaks \cite{noi}.
With the phase grid the range of existence of localized structures is enlarged from $I_{in}=0.32 \pm 0.03$ to $0.60 \pm 0.05$ $mW/cm^2$, since the pinning enhances the stability
of localized structures. The stability of the homogeneous solution is also increased, because the grid, that has a mismatched spatial frequency, damps the most unstable mode moving the bifurcation to a pattern state at higher input intensity.

The physical origin of the pinning mechanism has to searched in a small and local reorientation of the liquid crystal director in correspondence with the local
phase maxima of the input beam. In the experiment we do not have direct access to the molecular orientation angle $\theta$, but what we observe is the consequent localization of the light. This can be seen by considering the model, which
consists in a local relaxation equation for the average director tilt $\theta$, $0 < \theta < \pi/2$, coupled to the feedback write intensity $I_w$ \cite{19}. When the voltage across the liquid crystals is larger than the Fr\'eedericksz transition threshold \cite{11}, $V_{FT}$, the equation for $\theta$ reads as

\begin{equation}
\tau \partial _{t}\theta =l^{2}\nabla_\perp^2\theta -\theta + {\pi \over 2}\left(1- \sqrt{V_{FT} \over 
{\Gamma V_0 + \alpha I_w(\theta)}} \right)
\label{Eq1}
\end{equation}
where $l$ is the electric coherence length, $\tau$ the local relaxation time ($l=40$ $\mu m$, $\tau=30$ $ms$), 
$\Gamma V_0 + \alpha I_w$ is the effective voltage across the liquid crystals, with $\Gamma=0.3$ , $\alpha = 2.5$ $V,cm^2/mW$ , $V_0 = 12.8$ $ V$
and $V_{FT}=1.05$ $V$.
After a free propagation length $L$, the feedback light intensity is given by:

\begin{equation}
I_{w}= {I_{in} \over 4} \mid e^{i {L \lambda \over 4 \pi} \nabla_\perp^2} 
\cdot \left[ e^{i \varphi_{SPM}(x,y)} \left( 1- e^{- i \beta \cos^2 \theta} \right) \right] \mid^2 
\label{Eq2}
\end{equation}
the transverse Laplacian operator  accounting for diffraction and $I_{in}$ being the input light intensity.

\begin{figure} [h!]
\centerline{\epsfxsize=6 truecm \epsffile{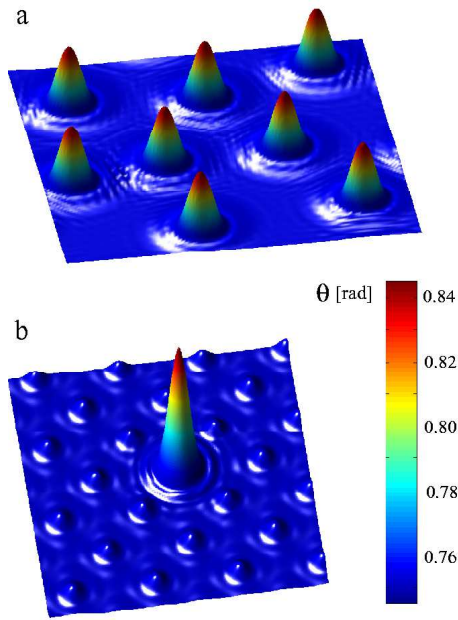}}
\caption{Numerical three-dimensional $\theta$ profiles; a) localized structures without the phase grid, b) a single localized structure pinned to the spatially modulated grid.
\label{fig3D}}
\end{figure}

\begin{figure} [h!]
\centerline{\epsfxsize=7 truecm \epsffile{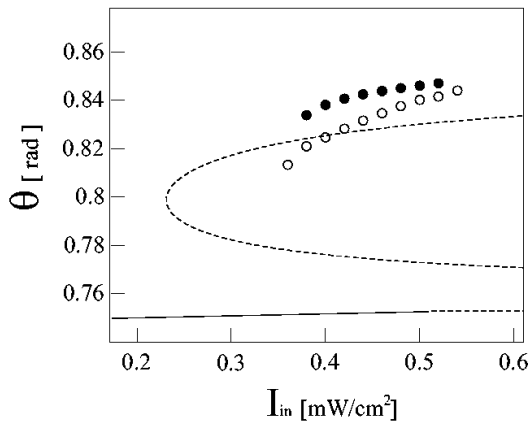}}
\caption{Bifurcation diagram as a function of $I_{in}$; solid/dashed lines are the results of the linear stability analysis (stable/unstable, respectively) of the homogeneous steady state; black empty/filled circles are the numerically calculated localized structure peak amplitudes with/without phase modulation.
\label{fig2}}
\end{figure}

Numerical simulations of the model equations Eqs.(\ref{Eq1}),(\ref{Eq2}) without the grid ($\varphi_{SPM}=0$), show the appearance of localized structures in the region of bistability between two differently oriented states. 
Three-dimensional $\theta$ profiles for this range of parameters
are displayed in Fig.\ref{fig3D}a. We see that adjacent localized structures interact through spatial oscillations in their tails \cite{16}. 
When the grid is introduced, the numerical $\theta$ profiles, displayed in Fig.\ref{fig3D}b, show a periodic modulation of the background, the amplitude of this 
modulation being less than $1 \%$ of the homogenous state and roughly  $7 \%$ of the localized structure peak amplitude. The local maxima of the grid act as pinning sites, leading to the suppression of the crosstalk.

The linear stability of the homogeneous steady state can be calculated analytically. This is shown by the solid/dashed lines in Fig.\ref{fig2}. When the phase grid is sent through the SPM, the bifurcation diagram is slightly modified
since the stability of the lower state is changed by the low intensity 
modulation introduced in the feedback loop. The black empty/filled circles in Fig.\ref{fig2}
are the numerically calculated peak amplitude of the localized structures with/without the phase grid, respectively. 
The bistable region is quantitatively consistent with the experimental observations, 
and with the enlarged 
region of existence of localized structures in the presence of the grid. Indeed, the pinning mechanism anticipates
the existence of localized structures and delays the instability of the lower homogeneous steady state, as observed experimentally.

In conclusion, we have shown that by introducing appropriate spatial modulation on the phase of the input beam, it is possible to suppress the crosstalk between adjacent localized structures, so that a large number of storable pixels can be put together. Each site on the matrix can be addressed by a local pulse or by images sent through a SPM. More generally, several control inputs could be coupled in such a way to act simultaneously, so that the demonstration here proposed could be easily extended to a higher level of parallelism. Such control capabilities might find application in optical tweezers and reconfigurable optical connections. 

U.B. acknowledges financial support from the FUNFACS European project, n¡. 2005-004/004868.

\end{document}